\documentclass{article}
\usepackage{spconf,amsmath,graphicx}
\usepackage{comment}
\usepackage{amssymb}
\usepackage{dsfont}
\usepackage{booktabs}
\usepackage{soul}
\usepackage{hyperref}
\usepackage{multirow}
\usepackage{xcolor}

\usepackage[hang,flushmargin]{footmisc} 
\usepackage{times} 
\usepackage{geometry} 

\renewcommand{\footnoterule}{%
  \kern -3pt
  \hrule width 1in
  \kern 2.6pt
}

\title{Subject Representation Learning from EEG using Graph Convolutional Variational Autoencoders}

\name{}
\address{Author Affiliation(s)}
\name{Aditya Mishra, Ahnaf Mozib Samin, Ali Etemad, Javad Hashemi
}
\address{Queen's University, Canada}
 
\begin{document}
%
\maketitle
\begin{abstract}
We propose GC-VASE, a graph convolutional-based variational autoencoder that leverages contrastive learning for subject representation learning from EEG data. Our method successfully learns robust subject-specific latent representations using the split-latent space architecture tailored for subject identification. To enhance the model's adaptability to unseen subjects without extensive retraining, we introduce an attention-based adapter network for fine-tuning, which reduces the computational cost of adapting the model to new subjects. Our method significantly outperforms other deep learning approaches, achieving state-of-the-art results with a subject balanced accuracy of 89.81\% on the ERP-Core dataset and 70.85\% on the SleepEDFx-20 dataset. After subject adaptive fine-tuning using adapters and attention layers, GC-VASE further improves the subject balanced accuracy to 90.31\% on ERP-Core. Additionally, we perform a detailed ablation study to highlight the impact of the key components of our method.
\end{abstract}


\section{Introduction}
\label{sec:introduction}
Representation learning in Electroencephalography (EEG) is a challenging yet essential task, with applications ranging from analyzing event-related potentials (ERPs) to gaining insights into cognitive processes and brain functions \cite{kappenman2021erp}. High noise and inter-subject variability in EEG data, complicate capturing generalizable latent representations.

Subject-specific variability is crucial for understanding individual differences in EEG signals, which could eventually serve as biomarkers. Studies \cite{hu2022new,khan2022contrastive} suggest that inter-subject variability should be regarded as a significant feature for understanding individual differences, rather than mere noise. Deep learning-based feature extraction has gained prominence in this domain, with autoencoders showing promise in learning feature-rich and transferable representations \cite{li2015feature,wen2018deep}. These representations are valuable for subject identification and adaptation to unseen conditions \cite{golmohammadi2019automatic,yin2017cross}. For a comprehensive review of deep learning-based EEG analysis, we refer the readers to \cite{roy2019deep}. However, we identify three key challenges in EEG feature extraction: \textbf{(i)} EEG data is characterized by a high degree of noise and variability across subjects, complicating the extraction of subject-specific latent representations. 
\textbf{(ii)} Inter-subject variability often interferes with recovering task-specific content, making it challenging to seperate subject-related variability (style) from task-related patterns (content). \textbf{(iii)} Achieving zero-shot generalization, allowing the model to adapt to unseen subjects without requiring additional data collection or training.

To address these challenges,
we propose GC-VASE, a \textbf{G}raph \textbf{C}onvolutional \textbf{V}ariational \textbf{A}utoencoder for \textbf{S}ubject representation learning from \textbf{E}EG. Our method effectively combines graph convolutional neural networks (GCNN) with variational autoencoders (VAE) to extract subject-specific latent representations from EEG data by splitting the latent space into subject latent space and residual latent space to improve subject identification. On the ERP-Core dataset, our model achieves 89.81\% zero-shot subject-balanced accuracy, rising to 90.31\% with fine-tuning. Further evaluation on the SleepEDFx-20 dataset \cite{kemp2000analysis} demonstrates its robustness and generalizability, achieving a subject-balanced accuracy of 70.85\%.

In summary, we make the following contributions: \textbf{(1)} We introduce a novel architecture for subject representation learning from EEG, which is based on graph neural networks (GNNs) and VAEs. 
\textbf{(2)} We employ subject-adaptive fine-tuning with adapters and attention layers to enhance the model's flexibility. This approach extends the model's applicability to a broader range of subjects, while minimizing the need for extensive retraining. 
\textbf{(3)} Experiments on the ERP-Core \cite{kappenman2021erp} and SleepEDFx-20 \cite{kemp2000analysis} datasets show that our approach outperforms previous methods, achieving state-of-the-art performance in subject identification.

\begin{figure*}
    \centering
    \includegraphics[width=0.95\textwidth]{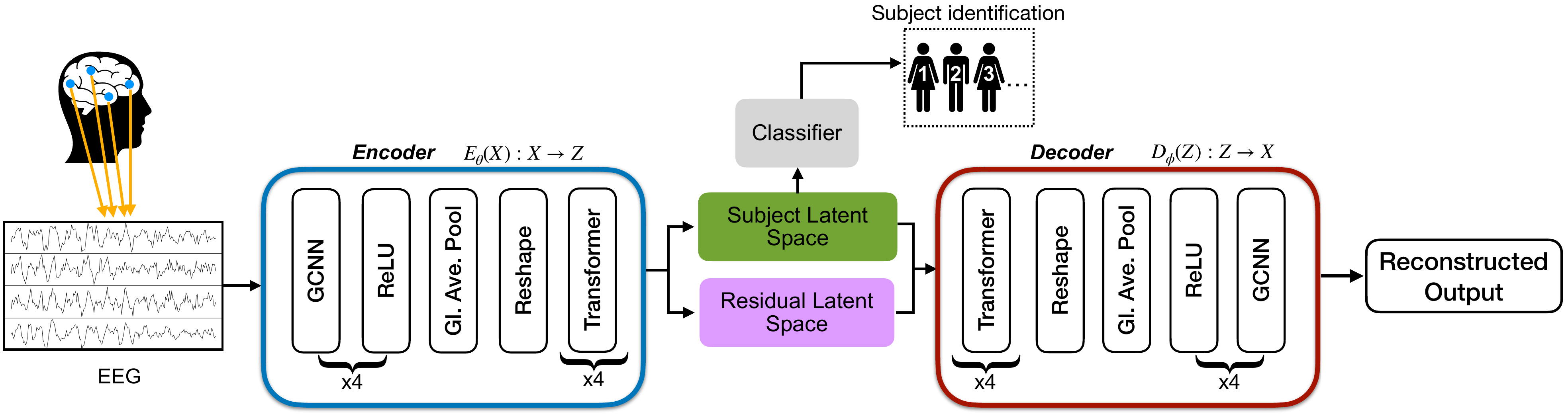}
    \caption{The proposed GC-VASE model incorporates a split latent space. The encoder splits the latent space into subject-specific space and residual latent space that are subsequently used for subject and task classification through an XGB classifier.}
    \label{fig:model}
\end{figure*}

\section{Related Work}
\label{sec:related_work}

Deep learning has advanced latent representation extraction for EEG-based subject identification.. Norskov et al. \cite{norskov2024cslp} proposed a VAE-CNN framework using contrastive learning to disentangle subject variability (style) from task activation (content). 
Wen et al. \cite{wen2018deep} introduced a deep convolution autoencoder for unsupervised EEG feature learning. Ng et al. \cite{ng2023deep} proposed a joint embedding VAE model for improved cross-subject generalization in EEG classification.

In related domains like speech processing and emotional recognition, separating subject-specific information has been shown to improve model generalization \cite{bollens2022learning, rayatdoost2021subject}. 
Rayatdoost et al. \cite{rayatdoost2021subject} used adversarial techniques to enhance emotion identification by minimizing subject-specific information. \cite{bollens2022learning} explored how explicitly modeling subject-invariant features can boost classification accuracy and model generalization across subjects, employing hierarchical VAEs to factor EEG data into two disentangled latent spaces.

To achieve local smoothness within classes, contrastive learning, a self-supervised learning technique is employed. 
Shen et al. \cite{shen2022contrastive} used contrastive learning to align EEG signal representations across subjects exposed to identical emotional stimuli. They used CNNs to learn spatio-temporal representations of EEG data that align across subjects. Chen et al. \cite{chen2020simple} introduced contrastive learning-based Self-supervised algorithms, focusing mainly on visual data. Mohsenvand et al. \cite{mohsenvand2020contrastive} extended the SimCLR framework for EEG to enhance sample efficiency by recombining multi-channel recordings.

GNNs have shown great results in the realm of representation learning as they also take into account the spatial representations of the EEG data. Behrouzi et al.\cite{behrouzi2021understanding} analyzed a GCNN-based EEG biometric system, capturing individual variability through graph-based functional relationships. Similarly, exploration of combinations of CNNs and GNNs to develop an end-to-end edge-aware spatio-temporal GCNN for EEG classification was used in \cite{li2019classify}. Zhong et al. \cite{zhong2020eeg} proposed a regularized graph neural network (RGNN) for emotion recognition, modeling inter-channel relations with adjacency matrices.

Jin et al. \cite{jin2024affective} proposed a multidomain coordinated attention transformer (MD-CAT), integrating spatial and time-frequency attention mechanisms for adaptive fine-grained feature extraction in affective EEG-based person identification in \cite{jin2024affective}. In \cite{moctezuma2019subjects}, EEG signals during imagined speech were explored as a biometric measure, achieving strong subject identification using wavelet and statistical features. Finally, a task-free biometric method using phase synchronization was introduced in \cite{kong2019eeg}, achieving high performance across multiple datasets.

\section{Proposed Method}
\label{sec:method}
The goal of our framework is to train a GCNN-based VAE with contrastive learning to accurately capture and disentangle subject-specific representations from residual latent representations in EEG data, enhancing subject identification. Figure \ref{fig:model} illustrates our model. Detailed explanations of the framework are provided later in this section.

Building on the potential of \textbf{VAEs} for learning feature-rich representations, we use a standard VAE model consisting of an encoder, denoted as $E_\theta(X) : X\rightarrow Z$, and a decoder, denoted as $D_\phi(Z) : Z\rightarrow X$. The encoder maps input data to a latent space, while the decoder reconstructs it from this latent representation. To capture spatial patterns, we integrate GNNs into both the VAE's encoder and decoder, inspired by recent advances in EEG representation learning. As illustrated in Figure \ref{fig:model}, the model consists of four GCNN layers with ReLU activation, followed by a global average pooling layer and a reshaping layer to prepare the data for subsequent transformer layers. To further capture complex temporal dependencies in EEG, both the encoder and decoder incorporate four-layer transformers on either side of the bottleneck.

Our method focuses on disentangling the latent space into subject-specific representations. The encoder explicitly splits the latent space, resulting in subject latents, $z^{S}$, and residual latents, $z^{T}$, represented as $E_\theta(X) = (z^{S},z^{T})$. However, since the focus of this work is on subject identification, we primarily utilize the subject-specific latent, $z^{S}$. The decoder reconstructs the input as $\hat{X} = D_\phi(z^{S},z^{T})$. Following \cite{norskov2024cslp}, we employ a shared encoder to minimize parameters, and we further pass the split latents into a extreme gradient boosting (XGB) classifier for subject identification.

To enhance the disentangling process, we integrate \textbf{contrastive learning}, which refines the subject representations by maximizing the similarity between positive pairs (same subject) and minimizing it between negative pairs (different subjects). This promotes more distinct subject representations in the latent space. We use contrastive learning on subject and residual split-latents by sampling pairs separately, which refines the latent space for either subject-specific or residual representations. We use multi-class N-pair loss \cite{sohn2016improved}, a deep metric learning method that constructs batches by sampling two examples from each class, forming K-pairs of samples. We combine this batch construction method with the InfoNCE generalization from \cite{oord2018representation}, incorporating the temperature scaling parameters $\tau$ from \cite{chen2020simple}. This is similar to the CLIP loss \cite{radford2021learning}, where we minimize the symmetric cross-entropy loss of the temperature-scaled similarity matrix, analogous to the NT-Xent loss \cite{chen2020simple}. 

Let $Z^{A} \in \mathbb{R^{C \times K}}$ and $Z^{B} \in \mathbb{R^{C \times K}}$ represent the latent embeddings of the K-pair samples. The loss $L_{NT-Xent}$ and $L_{CLIP}$ are defined by: 
\begin{equation}
\small
\label{eqn:nt-xent}
    L_{NT-Xent}(\mathcal{L}; Z', Z'', k)  = -log \frac{exp(\frac{sim(z'_{k},z''_{k})}{\tau})}{\sum_{i=1}^{k}\mathds{1}_{i \neq k}exp(\frac{sim(z'_{k},z''_{i})}{\tau})}
\end{equation}
 \begin{align}
 \small
 \label{eqn:clip}
    L_{CLIP}(\mathcal{L}; Z^{A}, Z^{B}) &= \frac{1}{K} \sum_{k=1}^{K} \bigg( L_{NT-Xent}(\mathcal{L}; Z^{A}, Z^{B}, k) \notag \\
    &\quad + L_{NT-Xent}(\mathcal{L}; Z^{B}, Z^{A}, k) \bigg)
 \end{align}
where $\mathds{1}_{[\textbf{c}]} \in {0,1}$ is an indicator function that returns 1 if condition \textbf{c} is true. $\mathcal{L}$ denotes the latent space from which the pairs have corresponding labels, and $sim(z'_{i}, z''_{k})$ represents the similarity metric. For the residual latent space $\mathcal{T}$ and subject latent space $\mathcal{S}$, $L_{CLIP}(\mathcal{T}; ., .,)$ and  $L_{CLIP}(\mathcal{S}; ., .,)$ correspond to the contrastive loss across residual tasks and subjects, respectively.

We further \textbf{fine-tune} the model using attention-based adapters that incorporate subject-specific attention while keeping the backbone network frozen. These adapters are lightweight neural modules added to the pre-trained GC-VASE's encoder. Unlike traditional fine-tuning, adapter-tuning reduces computational costs and overfitting risks. Moreover, it enhances the model's flexibility and adaptibility to new unseen subjects.

\section{Experiments}
\label{sec:experiments}
\noindent \textbf{Datasets.}
\label{ssec:data}
We conduct our experiments using the Event-Related Potentials-Core (\textbf{ERP-Core}) dataset \cite{kappenman2021erp}, consisting of 40 neurotypical young adults performing six distinct tasks. ERP-Core features seven commonly studied ERP components: (\textit{i}) N170 (Face Perception Paradigm), (\textit{ii}) MMN (Mismatch Negativity, Passive Auditory Oddball Paradigm), (\textit{iii}) N2pc (Simple Visual Search Paradigm), (\textit{iv}) N400 (Word Pair Judgement Paradigm), (\textit{v}) P3 (Active Visual Oddball Paradigm), (\textit{vi}) LRP and (\textit{vii}) ERN (Lateralized Readiness Potential and Error-related Negativity, Flankers Paradigm). 

We further evaluate our model using the \textbf{SleepEDFx-20} dataset \cite{kemp2000analysis}, which includes 197 whole-night PolySomnoGraphic sleep recordings with EEG, EOG, chin EMG, and event markers. To address the limited number of EEG channels, we focus on a single channel and apply a short-time Fourier transform to align the SleepEDFx-20 data with the configuration used for ERP-Core.

\noindent \textbf{Implementation details.}
\label{ssec:implement}
The experiments utilize raw time-domain EEG signals sampled at 256 Hz. The length of an epoch window is 1-second long with a time resolution of 256 samples. The EEG signals underwent low-pass filtering using a 5th-order sinc filter (cutoff: 204.8 Hz) followed by a high-pass filtering with a non-causal Butterworth filter (cutoff: 0.1 Hz, roll-off: 12 dB/octave). Following \cite{norskov2024cslp}, we use a 70:10:20 ratio for training, development, and testing on the ERP-Core dataset. Additionally, we employ a 30-second time windowing method as described in \cite{norskov2024cslp} for the SleepEDFx dataset. Consistent with \cite{norskov2024cslp}, we adopt a 5-fold cross-validation approach, training a new model with the same architecture for each fold.

We employ L4 and A100 GPUs with a batch size of 256. The models are trained using the Adam Optimizer with a learning rate of $1\times10^{-4}$ over 200 epochs. The latent space dimentionality is set to 64, and the softmax temperature ($\tau$) is initialized at 14.29. To obtain final results, we average the scores across three seed values. Similar to the training process, we use a kernel of size 4 and input channels of 30 for evaluation. A 5-fold cross-validation scheme is used, followed by zero shot inference.

For subject-adaptive fine-tuning, we add an adapter, consisting of two layers of a multi-headed attention mechanism, to the end of the model's encoder. Fine-tuning is performed over 20 epochs with a batch size of 256, and the number of heads in the attention layers is set to 8. The FT variant is fine-tuned on 70\% of each test participant's data and evaluated on the remaining 30\% of unseen data.

\begin{figure}[t]
    \centering
    \includegraphics[width=\columnwidth]{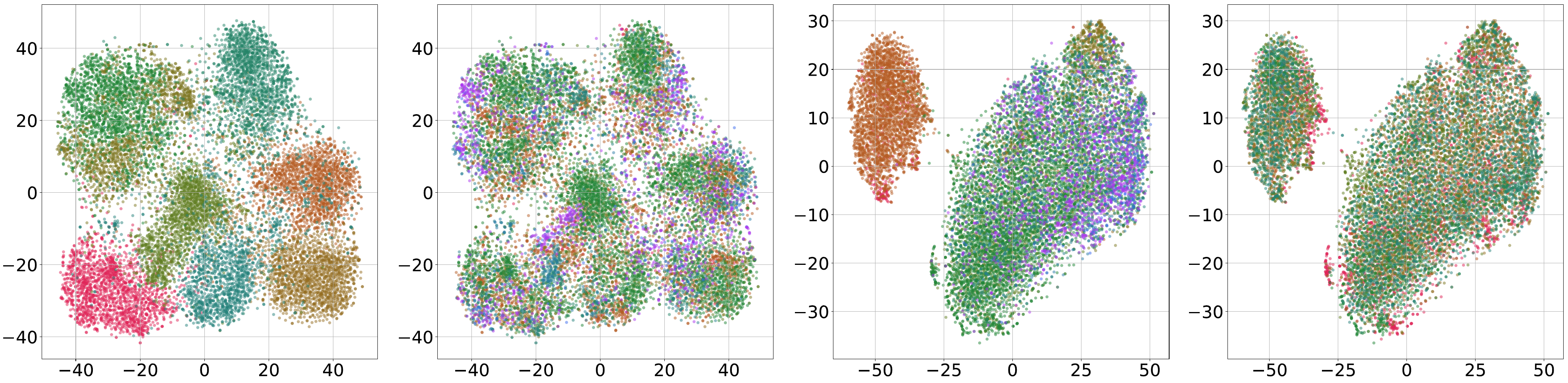}
    \caption{t-SNE plots of split-latents encoded on the test set (unseen subjetcs), colored by their true labels. The plots, in order, depict: subject space colored by subject, subject space colored by task, task space colored by task, and task space colored by subject.}
    \label{fig:tsne}
\end{figure}

\noindent \textbf{Results.}
\label{ssec:results}
In line with previous work \cite{norskov2024cslp}, we evaluate GC-VASE on the ERP-Core benchmark using 5-fold cross-validation. The results for subject identification and task classification are summarized in Table \ref{table-0}. All results in Table \ref{table-0} are computed using a zero-shot approach. The results demonstrate that GC-VASE surpasses the prior state-of-the-art approaches \cite{norskov2024cslp} achieving up to 9.49\% improvement in subject identification balanced accuracy. Additionally, the competitive performance on task classification using residual latent space highlights the potential adaptability of GC-VASE to other EEG-related tasks.

\begin{table}[t]
\centering
\caption{Subject identification and task classification balanced accuracies (\%) on the ERP-Core dataset.}
\label{table-0}
\setlength{\tabcolsep}{2pt}
\resizebox{0.95\columnwidth}{!}{
\begin{tabular}{lcccc}
\toprule
\multirow{2}{*}{\textbf{Model}} & \multicolumn{2}{c}{\textbf{Subject Identification}} & \multicolumn{2}{c}{\textbf{Task classification}}\\
 & \textbf{Subject latent} & \textbf{Residual latent} & \textbf{Residual latent} & \textbf{Subject latent}\\
\midrule
CSLP-AE 
& 80.32 & 79.64 & \textbf{48.48} & 45.41\\
SLP-AE 
& 74.63 & 74.70 & 47.00 & 47.23\\
C-AE 
& 79.42 & 73.27 & 46.59 & 37.34\\
AE 
& 60.68 & 61.08 & 31.43 & 31.62\\
GC-VASE & \textbf{89.81} & 85.40 & 36.18 & 31.83\\
\bottomrule
\end{tabular}
}
\end{table}

\begin{table}[t]
\centering
\caption{Ablation study on the impact of major components in GC-VASE for subject identification on ERP-Core.} 
\label{table:ablation}
\setlength{\tabcolsep}{10pt}
\resizebox{0.9\columnwidth}{!}{
\begin{tabular}{lcc}
\toprule
\textbf{Variants} & \textbf{F1-score (\%)} & \textbf{Accuracy (\%)}\\
\midrule
GC-VASE & \textbf{89.58} & \textbf{89.81} \\ 
w/o GCNN & 81.04 ($\downarrow$ 8.54) & 81.73 ($\downarrow$ 8.08)\\ 
w/o contrastive learning & 80.18 ($\downarrow$ 9.40) & 80.84 ($\downarrow$ 8.97)\\
w/o split-latent & 81.51 ($\downarrow$ 8.07) & 81.72 ($\downarrow$ 8.02)\\
\bottomrule
\end{tabular}
}
\end{table}

\begin{table}[t]
\centering
\caption{GC-VASE subject identification performance across various paradigms of ERP-Core dataset.}
\label{table:paradigm_wise}
\setlength{\tabcolsep}{10pt}
\resizebox{0.6\columnwidth}{!}{
\begin{tabular}{lll}
\toprule
\textbf{Paradigms} & \textbf{Balanced} & \textbf{Closed-Set} \\
\midrule
N400 & 98.87 & 99.12 \\ 
P3 & 57.67 & 83.23 \\ 
ERN & 59.89 & 79.52 \\ 
N2pc & 50.49 & 79.56 \\ 
N170 & 36.32 & 65.78 \\ 
MMN & 41.03 & 64.04 \\ 
\bottomrule
Average & 57.37 & 78.54 \\ 
\bottomrule
\end{tabular}
}
\end{table}

\begin{table}[t]
\centering
\caption{Subject identification and task classification balanced accuracies (\%) on SleepEDFx-20 with 20 subjects.}
\label{table:sleep}
\setlength{\tabcolsep}{3pt}
\resizebox{1\columnwidth}{!}{
\begin{tabular}{lccccc}
\toprule
\multirow{2}{*}{\textbf{Model}} & \multirow{2}{*}{\textbf{Split Latent}} & \multicolumn{2}{c}{\textbf{Subject Identification}} & \multicolumn{2}{c}{\textbf{Task classification}}\\
 & & \textbf{Sub. latent} & \textbf{Res. latent} & \textbf{Res. latent} & \textbf{Sub. latent}\\
\midrule
CNN 
& No & 67.18 &  - & - & -\\
LaBraM \cite{jiang2024large}
& No & 59.42 & - & - & -\\
CSLP-AE \cite{norskov2024cslp} 
& Yes& 67.55 & 67.18 & 34.91 & 34.48 \\
GC-VASE & Yes & \textbf{70.85} &  70.60 & \textbf{46.19}  & 45.91\\
\bottomrule
\end{tabular}
}
\end{table}

An interesting pattern emerges from our experiment: more convolutional layers in the encoder improve task classification, while more GCNN layers enhance subject identification. If GC-VASE's encoder and decoder use only convolutional layers (as in \cite{norskov2024cslp}), task accuracy increases, but subject accuracy declines. Convolutional layers capture spatial hierarchies and local patterns essential for task classification. Adding more layers enables the model to extract higher-level, task-specific features. Conversely, GCNNs are particularly effective at capturing relational or graph-structured data, such as inter-subject relationships. We also evaluate subject identification and task classification performance using representations from the residual and subject latent space, respectively. Here, GC-VASE is trained on one latent representation to predict the other. The model maintains disentangled latent spaces while preserving the structural encoding necessary for accurate subject and task classification. The latent-permutation method replaces the standard autoencoder reconstruction loss and is used with contrastive learning and batch construction to maintain the disentangled latent spaces. 
The strong performance on ERP-Core demonstrates the effectiveness of the split-latent space approach for extracting subject representation. After fine-tuning with adapters for subject adaptation, the balanced accuracy of GC-VASE reaches 90.31\%.

An \textbf{ablation study} highlights the importance of key components in GC-VASE, as shown in Table \ref{table:ablation}. Removing the GCNN layers from the encoder and decoder results in an 8.54\% drop in subject-balanced accuracy, while removing contrastive learning causes a 9.40\% decline. A similar performance decline is observed when the split-latent space is removed and replaced with a single latent space that contains both subject and residual representations. We apply split-latent permutation loss after removing the contrastive loss. These significant drops emphasize the critical role of GCNN layers and contrastive learning in subject identification. Fewer GCNN layers lead to poor encoding of inter-subject variability, making it difficult for the model to disentangle subject latent representations, thereby reducing overall performance. However, there is a trade-off between the performance and computational cost as we add more GCNN layers. Upon replacing the VAE with an AE in GC-VASE, we observe a decline in performance, with subject accuracy dropping to 85.31\% (↓ 4.5\%). To better understand the learned latent representations, we visualize the split-latent space using t-SNE (Figure \ref{fig:tsne}). In the subject space, distinct clusters by subject suggest effective capture of subject-specific features, consistent with the high classification accuracy in Table \ref{table-0}. When colored by task, these clusters remain subject-organized, indicating a focus on subject identity. In the residual space, tasks are well-separated, with subjects intermixed. Table \ref{table:paradigm_wise} provides ERP-Core's paradigm-wise identification results, showing both balanced and closed-set accuracy. We can observe noticeable performance differences across various paradigms.  

Table \ref{table:sleep} demonstrates the performance of GC-VASE on the SleepEDFx dataset, where GC-VASE successfully outperforms other methods. We observe that our method achieves a 3.3\% improvement in subject classification and an 11.28\% improvement in task classification compared to CSLP-AE \cite{norskov2024cslp}. Similarly, our method outperforms LaBram \cite{jiang2024large} by a significant margin of 11.43\%.

\section{Conclusion}
\label{sec:conclusion}

We introduce GC-VASE, a novel VAE based on GCNNs that leverages contrastive learning for disentangling subject-specific features. Our approach effectively isolates subject latents, resulting in enhanced performance for subject identification. Experiments on the widely used ERP-Core benchmark and SleepEDFx-20 dataset demonstrate that our method significantly outperforms existing state-of-the-art techniques, achieving a subject-balanced accuracy of 89.81\% and 70.85\%, respectively. Additionally, the lightweight, fine-tunable adapters integrated into GC-VASE provide a flexible solution for scenarios with limited computational resources, enabling domain adaptation without the need for extensive retraining. Our ablation study demonstrates the importance of using GCNN and contrastive learning in GC-VASE. This research offers promising practical applications in areas such as personalized medical diagnostics and EEG-based biometric systems. Our future work includes leveraging knowledge distillation in GC-VASE from the large EEG foundation models for subject identification.

\noindent \textbf{Acknowledgement.}
This work was supported by Mitacs Globalink Research Internship (GRI) program. 



\small
\bibliographystyle{IEEEbib}
\bibliography{refs}

\end{document}